\documentclass[onecolumn,12pt,pre]{revtex4}
\usepackage{subfig}
\usepackage[final]{graphicx}
\usepackage{amsmath}
\usepackage{amssymb}
\usepackage{amscd}
\usepackage{color}
\usepackage{bm}

\begin{document}

\title{Controlling the composition of a confined fluid by an electric field}
\author{C. Brunet, J. G. Malherbe and S. Amokrane}
\affiliation{Physique des Liquides et Milieux Complexes, Facult\'{e} des Sciences et de
Technologie,Universit\'{e} Paris Est, 61 av. du G\'{e}n\'{e}ral de Gaulle,
94010 Cr\'{e}teil Cedex, France}
\date{18 August 2009}

\begin{abstract}
Starting from a generic model of a pore/bulk mixture equilibrium, we propose
a novel method for modulating the composition of the confined fluid without
having to modify the bulk state. To achieve this, two basic mechanisms -
sensitivity of the pore filling to the bulk thermodynamic state and electric
field effect - are combined. We show by Monte Carlo simulation that the
composition can be controlled both in a continuous and in a jumpwise way.
Near the bulk demixing instability, we demonstrate a field induced
population inversion in the pore. The conditions for the realization of this
method should be best met with colloids, but being based on robust and
generic mechanisms, it should also be applicable to some molecular fluids.
\end{abstract}

\maketitle

\section{Introduction}
The question of how the adsorption of foreign particles affects the
properties of materials and the means to control this is of central
importance in domains ranging from separation processes to nanotechnology.
This motivates the continuing investigation on the factors determining the
adsorption process and the search of conditions most favorable for its
control (composition of the adsorbing fluid, adsorption geometry etc.) \cite%
{Fundamentals_Adsorp}. The purpose of this communication is to propose a
novel method that allows this, on the basis of Monte Carlo simulations of a
fluid-slit pore equilibrium. To avoid having to consider specific
interactions as in molecular adsorption, we choose to point out the basic
mechanisms on a simple model with only hard-sphere and dipolar interactions.
The situation closest to this model is then the adsorption of
macroparticles. Another reason is the recent development of studies of
colloidal adsorption \cite{Bechinger}. Indeed, while numerous studies exist
on molecular adsorption (see for e.g. \cite{Gubbins,LachetFuchs} and refs 
\cite{ Select_ads} for more recent work), one practical advantage of using
colloids is the possibility to tune their effective interaction (eg. by
adding polymeric depletants) and their coupling with external fields,
possibly in confined geometry \cite{VanBlaad}. The actual behavior may,
however, be complicated by the interplay of different effects (see for
example the role of static and hydrodynamic forces in microfluidics devices 
\cite{Ajdari}). We thus propose here a method that allows a fine control of
the composition of the adsorbed fluid, while remaining very simple.

\section{Method}
We start with the simplest confinement geometry: an open slit pore with
parallel walls in equilibrium with a bulk fluid. It has been used in several
theoretical studies to determine the parameters affecting the behavior of
the confined fluid, (see for example \cite%
{Gubbins,Sarkisov,Duda1,JCPAbd,JPCGub,Virgiliis,Duda2,Kim} and references
therein). Since we seek a method in which the pore geometry, the
interactions (between the particles and the particles and the confining
medium) as well as the bulk thermodynamic state are fixed, one alternative
is the coupling with an external field. This should always be possible since
besides particles having a permanent dipole such as magnetic colloids,
colloidal particles are always polarizable to some extent. We thus took a
uniform electric field $\bm{E}=E\ \bm{u_z}$ normal to the walls. As in \cite%
{JCPCharles} we considered a mixture in which one species bears a dipole
moment $\bm{\mu}$, taken permanent for simplicity. The field is then not
applied in the bulk. We thus have pure hard-spheres (species 1) and dipolar
hard-spheres (species 2), possibly with a non-additive diameter$\ \sigma
_{12}$ in the potential $u_{12}^{HS}$ (see also the discussion of figure 5).
Both species\ have a hard-sphere interactions with the walls. This makes the
model more appropriate \cite{JPCGub} to a mixture of hard-sphere-like
colloidal particles, than to a molecular mixture (see however the final
remarks). The effect of an external field (and temperature) on the filling
of a cylindrical pore was also studied in \cite{Rasaiah2} (see also and \cite%
{Bratko} for a slit pore), but not from a bulk mixture. Previous studies
considered the role of the pressure in one-component fluids (eg. \cite%
{Sarkisov}), or the total density and the mole fractions \cite%
{Duda1,Duda2,Kim} in bulk mixtures but without field. As shown below, the
combination of both will play here a crucial role. An inhomogeneous
multicomponent mixture with anisotropic interactions being difficult to
study by analytical methods (we are aware of one study by density functional
theory of the adsorption from a mixture of polar molecules \cite{Kotdawala}%
), we used Monte Carlo simulation (see also, for e.g. \cite%
{Rasaiah,Klapp,Rasaiah2,Weis,JCPCharles}). We already pointed out how the
structure can be modulated by the combination of various interactions \cite%
{JCPAbd,JPCGub} and by the action of the field \cite{JCPCharles}. However,
only the density profile of the particles through the pore could be
modulated in \cite{JCPCharles} since the total number of particles was kept
fixed (simulations in the canonical ensemble). An important difference here
is that we consider an open pore which exchanges particles with a reservoir.
One may then achieve much stronger variations of the density of each
component in the pore. The physical pore is assumed large enough that the
interfacial region in which it is in contact with the reservoir \ plays a
negligible role \cite{Panagiotop}.\ For this reason we will refer to the
fluid in the reservoir far from this region as the "bulk". The pore/fluid
equilibrium is determined by the equality of the chemical potentials $\mu
_{1}$ and $\mu _{2}$ of both species in the bulk and in the pore. But since
the practical control variables are the total density $\rho _{b}$ and mole
fraction $x_{2}$ of the dipolar species, in the bulk, the latter is studied
in the canonical ensemble. By considering only homogeneous states or
metastable states very close to the coexistence boundary, $\mu _{1}$ and $%
\mu _{2}$ in the bulk are determined with sufficient accuracy from Widom's
insertion method (see for eg \cite{Neimark} for this point). $\mu _{1}$ and $%
\mu _{2}$ are then used to study the fluid in the pore in the
grand-canonical ensemble. We can then compute the average density of each
species in the pore as a function of $\rho _{b}$ and $x_{2}$. Hereafter,
reduced variables $E^{\ast }=E(\sigma ^{3}/kT)^{1/2}$ and $\mu ^{\ast }=\mu
/(kT\sigma ^{3})^{1/2}$ will be used. The reduced density in the pore is $%
\rho =\bar{N}\sigma ^{3}/V$, with $\bar{N}$ the average number of particles
for a lateral surface $S$ with periodic conditions in the $x$ and $y$
directions and $V=S(H-\sigma )$ the accessible volume in the pore. We took a
pore width $H=3\sigma $. In the bulk, $N=N_{1}+N_{2}$ is fixed.

\section{Results}
We show in figure 1 the first basic mechanism: a field induced filling of
the pore by a one-component dipolar fluid. At increasing field strength $%
E^{\ast }$, the pore is progressively filled by the dipoles with a rate that
depends on $\rho_{b}$ . For the value of the $E^{\ast}$ and $\mu ^{\ast}$
used here, the explanation seems that the field-dipole interaction energy $%
-\sum_{i}\bm{\mu_{i}}.\bm{E}$ offsets the entropy loss due to their
orientation in the direction of the field. This is more visible at low $\rho
_{b}$ in which case the slope is nearly constant beyond $E^{\ast }=8$. For a
particle diameter of $1\mu m$ and $T=300K$, for example, this corresponds to 
$E=49\ 10^{-3}V/\mu m$ and $\mu =2\ 10^{5}D$. Thanks to the scaling factor $%
\sigma ^{-3/2}$ in its definition, the same value of $\mu^{\ast }$ is also
appropriate for the dipolar interaction between \textit{molecular} species.
\begin{figure}[htbp]
\centering
\includegraphics[angle=270,totalheight=7cm]{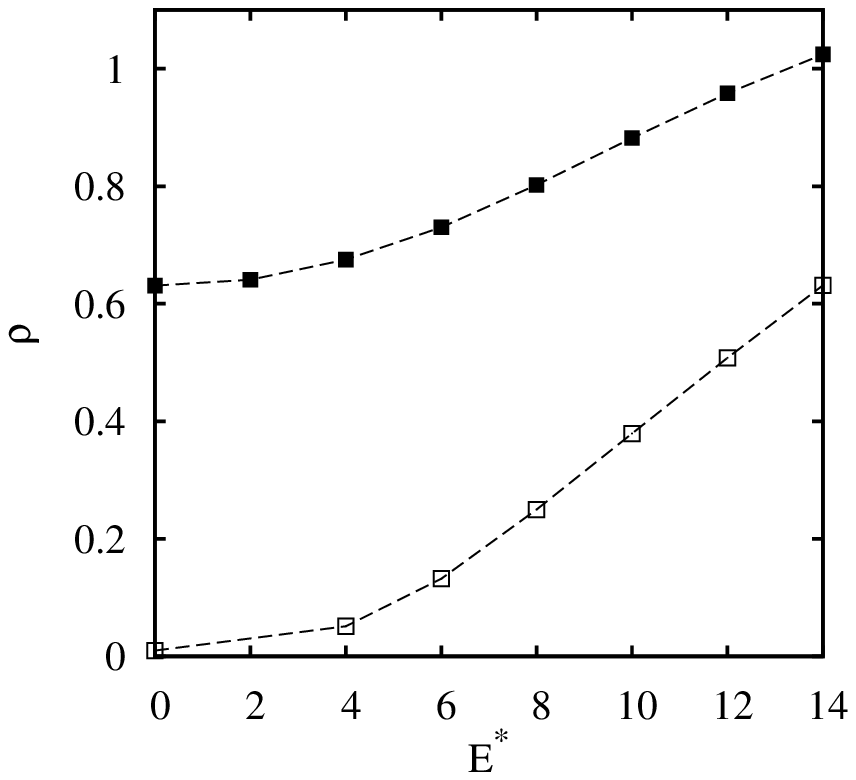}
\caption{Effect of the applied field on the filling of the pore by a
one-component dipolar fluid.\newline
$\rho $ is the total density in the pore and $E^*$ the field strength in
reduced units. The bulk density is $\rho_b=0.51$ (\emph{filled squares}) and $\rho_b=0.0102 $ (\emph{empty squares}). The lines are a guide to the eye.}
   \label{f:pore}
\end{figure}

The dipole moment being then of the order of one Debye, a field strength of
the order $10V/nm$ is needed to obtain the same reduced energy $-\mu ^{\ast
}E^{\ast }$. Such field strengths are not unusual for confinement at the
molecular scale (for water in nanopores see for example \cite{Rasaiah2} and 
\cite{Bratko}, in particular figure 1 in the last one).\ Note that the
equilibrium state in presence of the field may not always be the filled one
at other parameters or if more complex interactions are considered (see ref 
\cite{Rasaiah2} for example), due to the competition between energetic and
entropic contributions.

Figure 2 illustrates the second mechanism: the relative population of a pore
in equilibrium with a mixture having a natural tendency to demix. The
simplest one is the mixture of non-additive hard spheres \cite{amar} in
which the cross diameter is $\sigma_{ij}=1/2(\sigma_i+\sigma_j)(1+\delta).$
Previous studies \cite{Duda1,Duda2,Kim} have shown that when the pore is in
equilibrium with a mixture in which one species is in minority (say $%
x_2=0.02 $ for a non additivity parameter $\delta =0.2$ ) a population
inversion occurs in the pore when the total bulk density is varied. This
occurs here for $\rho_b$ between $0.55$ and $0.56$ for a pure non-additive
HS mixture and between $0.54$ and $0.55$ for the hard-sphere dipole mixture.
\begin{figure}[htbp]
\begin{center}

\subfloat[ Mixture of symmetric non-additive hard-spheres. \emph{Filled
circles}: adsorption; \emph{Open circles} : desorption; the bulk
concentration of the adsorbing species is $x_2=0.02$. The non-additivity
parameter is $\delta=0.2$
 ] {\label{f:fig2a}\includegraphics[angle=270,origin=br,totalheight=7cm]{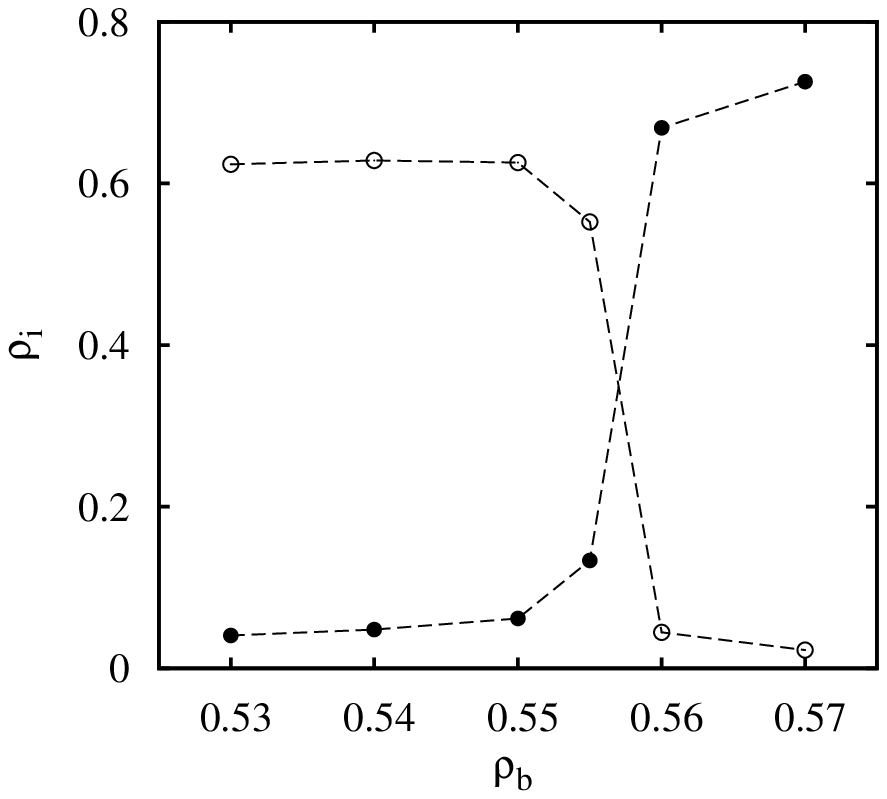}}
\subfloat[ Mixture of hard-spheres and dipolar hard-spheres with $\delta=0.2$%
, $\mu^*=1$ and $x_2=0.02$ ; \emph{filled circles:} adsorption of the
dipolar hard-spheres.]{\label{f:fig2b}\includegraphics[angle=270,origin=br,totalheight=7cm]{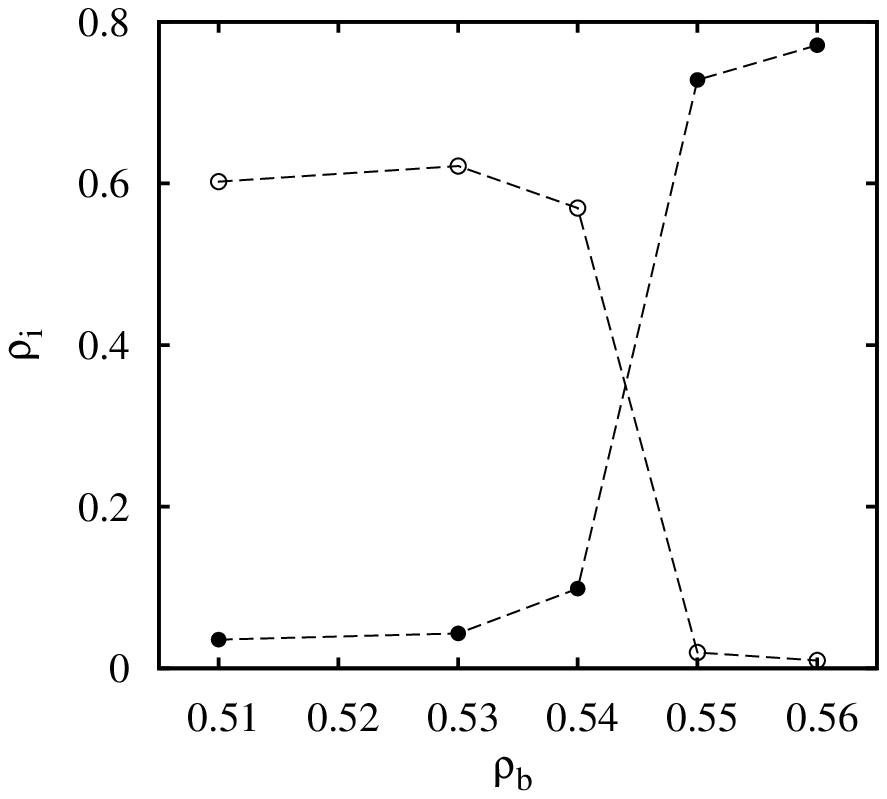}}
\caption{Population inversion in a pore in equilibrium with a bulk
mixture close to demixing.  }
\label{f:2}
\end{center}
\end{figure}

Having the basic ingredients, we may now combine them to produce the desired
effect: by choosing the composition of the bulk fluid so as to be close to
the population inversion in the pore, we anticipate that the closer we are
from the threshold density the weaker will be the external field $E^*_{tr}$
required to trigger it. This is shown in figure 3. In the most favorable
case shown the actual value of $E_{tr}$ is about $3 \ 10^{-3} V/\mu m$.

\begin{figure}[htbp]
\begin{center}
\includegraphics[angle=270,totalheight=7cm]{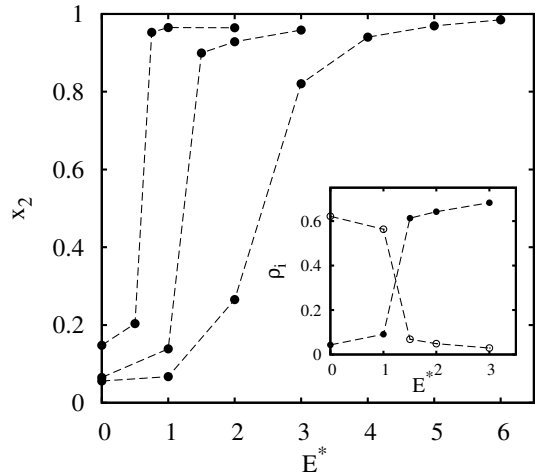}

\caption{The curves show the dipoles mole fraction in the pore as a function of the
reduced applied field strength. The curves are for a bulk mole fraction $%
x_2=0.02$ and bulk densities (from right to left) $\rho_b=0.51, 0.53, 0.54 $%
. The inset shows the corresponding dipoles and hard-spheres density in the
pore for $\rho_b= 0.53 $. }
   \label{f:fig3}
\end{center}
\end{figure}
A small variation of the applied field produces the inversion: the dipolar
particles are selectively absorbed by a weak change about $E_{tr}$, the
converse being possible, perhaps with some hysteresis \cite{Duda2}. Near the
adorption jump ($\rho _{b}=0.53$, $E^{\ast }=0.5$ for $\mu ^{\ast }=1$) a
slight change in temperature ($\delta T=25^{o}C$ in figure 4) produces a
detectable change in adsorption. 
\begin{figure}[htbp]
\begin{center}
\includegraphics[angle=270,totalheight=7cm]{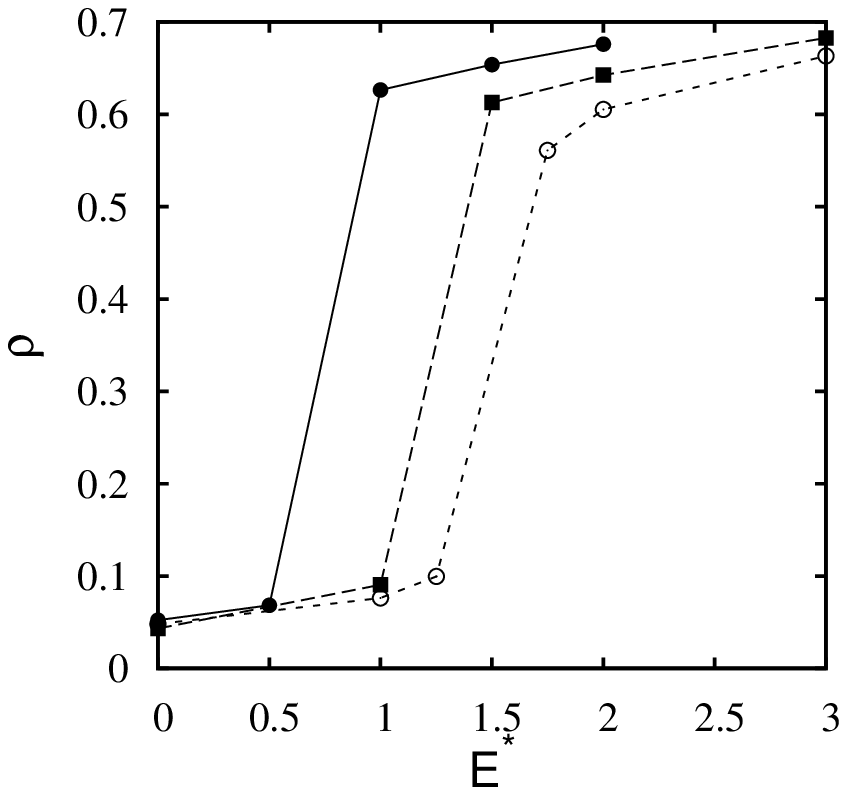}

\caption{ Adsorption-field strength curves at different temperatures.\newline
Dipoles mean density in the pore as a function of the
applied field strength at $T=285K$, $300K$ and $325K$ from left to right.
The bulk mole fraction is $x_{2}=0.02$ and bulk densities
are $\rho_b=0.53$. Here $E^*(T)=E^*(300)(T/300)^{1/2}$. }
   \label{f:fig4}
\end{center}
\end{figure}
This observation may be important for some
applications (since the reduced variables combine $E$ and $\mu $ with T,
while $E=0$ in the bulk, one has to rescale $E^{\ast }$ by the factor $%
T^{1/2}$ to compare two temperatures at a given values of $E$). 

Finally, in order to check the effect of a different interaction $u_{12}$
between species 1 and 2, we also show in figure 5 the result for a Yukawa
repulsive potential.
\begin{figure}[htbp]
\begin{center}
\includegraphics[angle=270,totalheight=7cm]{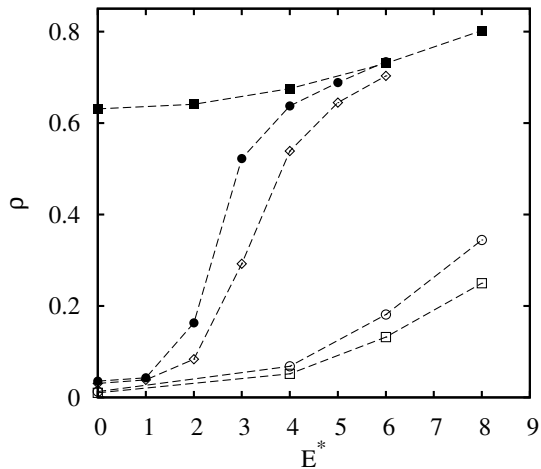}

\caption{Effect of the applied field on the filling of the pore by
different model fluids.\newline
$\rho $ is the density of the dipolar species, in the pore. \emph{Empty
squares} : one-component dipolar fluid (for $\rho _{b}=0.0102$ as in figure
1); \emph{open circles} : additive mixture of hard spheres and dipoles (with 
$x_{2}=0.02$ and $\rho _{b}=0.51$ in the bulk). \emph{Diamonds} : same for a
Yukawa repulsion between the dipoles and the hard-spheres; \emph{filled
circles} : same for the non-additive hard-spheres - dipolar hard-spheres
mixture. The range of the Yukawa potential (with $\epsilon \ast =8$) gives
the same contribution to the second virial coefficient as the non-additive
hard-spheres potential with $\delta =0.2$. \emph{Filled squares} :
one-component dipolar fluid with $\rho _{b}=0.51$ (as in figure 1).  }
   \label{f:fig5}
\end{center}
\end{figure}

  We observe that the phenomenon is quite general. The
requirement for observing a sensitive field effect is that the self
coordination should be more favored in the mixture. The poor miscibility can
be favored by suitable chemical composition of the particles surface layers \cite%
{Pusey,Hennequin}. Weak specific interaction with the
pore walls can be achieved similarly.

As our main goal was to demonstrate the phenomenon of a field activated
adsorption from an unstable mixture, we used a simple simulation strategy.
Accordingly, we did not conduct a detailed study of the behavior of the
confined fluid. For instance, phase equilibrium in the pore may take place
before the spontaneous condensation (adsorption jump) predicted in the grand
canonical simulation \cite{NeimarkPRE}. According to Duda al.\cite{Duda2},
the inversion line for non-additive hard spheres is close to the bulk fluid
coexistence line but the two phenomena are different. We actually observed
that the inversion corresponds to a bulk fluid close to demixing or slightly
in the two-phase region. Regardless of this, the essential point is that the
density in the liquid-like phase should be close to the value after the
adsorption jump, as in one-component systems \cite{NeimarkPRE}. The precise
relation between these observations and other phenomena such as capillary
condensation, wetting, hysteresis, etc. (see e.g. \cite%
{Wetting,NeimarkPRE,Kierlik}) will be discussed in future work.

\section{Conclusion}
In conclusion, these results show that the combination of two generic
mechanisms allows a quite sensitive control of the pore filling. Although
this method has been demonstrated for particles that are closest to the
optimum conditions (ie hard-sphere-like colloids), none of them is exclusive
and since the basic mechanisms (demixing instability and coupling with an
external field) are quite generic, this prediction should concern a broader
class of systems (including molecular ones). In order to benefit from the
field effect, one species should be either polar (e.g. ferrocolloid in
magnetic fields) or much more polarizable than the other (the results given
here being relative to permanent dipoles). The solution should also not
contain free charges to avoid particle motion due to the action of the field
(electrohydrodynamic flows), not considered in this simple model. Polar
molecules being on the other hand rather common, one should consider in this
case also the role of specific interactions. We believe that further
experimental studies and simulations on this method are worthwhile given the
diversity of possible applications of this field controlled composition of
the confined fluid and hence flexible control of the physical properties
that depend on the composition of the confined fluid. Just as an example,
one may consider to modulate in this way the dielectric response of the
confined fluid for optical applications.

\end{document}